# Physics-Based versus Data-Driven Classification of Single-Photon Quantum Emitters from Sparse Autocorrelation Data

*Nhat Minh Nguyen,[1] Md Shakhawath Hossain,[1] Duc Anh Ngo,[1] Chaohao Chen,[1] Xiaoxue Xu,[1] Toan Trong Tran[1,*] and Carlo Bradac[2,*]*

[1] School of Electrical, Mechanical and Biomedical Engineering, University of Technology Sydney, Ultimo, NSW, 2007, Australia

[2] Department of Physics & Astronomy, Trent University, 1600 West Bank Dr., Peterborough, Ontario K9L 0G2, Canada

*Corresponding author, e-mail: carlobradac@trentu.ca
*Corresponding author, e-mail: trongtoan.tran@uts.edu.au

**Abstract**
Identifying single-photon emitters from large, inhomogeneous candidate populations is key to realizing many quantum applications. This requires measuring the emitters' second-order autocorrelation function, whose statistical reliability is fundamentally limited by acquisition time. Machine-learning classifiers can accelerate identification from sparse data, but their performance relative to sequential, physics-based inference has not been systematically examined. Here, we introduce sequential Bayesian inference for single-photon-emitter classification and benchmark it against Levenberg–Marquardt fitting and a feedforward neural network. We use synthetic training and test data calibrated against real Hanbury Brown–Twiss measurements from hexagonal boron nitride emitters, enabling evaluation against an exactly known ground-truth emitter number under realistic noise and background conditions. All three approaches ultimately achieve high, near-perfect accuracy with sufficient integration time, but differ substantially in convergence rate and robustness under sparse photon statistics. The neural network is most robust at short integration times. The Bayesian classifier reaches stable, near-perfect accuracy fastest, once sufficient evidence has accumulated, while retaining full physical interpretability. Levenberg–Marquardt fitting remains a valuable, fully interpretable method, ultimately achieving the highest recall despite being the slowest to converge. These results lead to several key conclusions. No single method dominates across all performance metrics. Relying on any one metric alone can give a misleading picture of classifier performance, particularly under sparse photon statistics. Physics-based and data-driven methods are complementary rather than competing approaches. Combining all three predictions through a simple majority vote further improves overall classification performance, reflecting their partially independent failure modes. Together, these findings provide practical guidance for selecting and combining classification strategies for scalable single-photon-source screening and other quantum-emitter characterization tasks.

**Keywords:** single-photon sources, second-order autocorrelation, machine learning, Bayesian inference, hexagonal boron nitride, quantum emitter classification.

## 1. Introduction

Single-photon emitters (SPEs) are the fundamental constituents of many quantum-based applications and technologies. They encompass a wide range of physical systems, including semiconductor quantum dots,[1] point defects in solid-state materials,[2–8] as well as individual atoms,[9] ions,[10] and molecules.[11] Owing to their ability to generate single photons on demand, SPEs have become key components in numerous quantum technologies, including quantum key distribution, networking and communication,[12–15] as well as quantum computing,[16,17] quantum simulation,[18] and quantum sensing and metrology.[19–21] However, realizing these applications at scale requires the ability to rapidly and reliably identify single-photon emitters from large, inhomogeneous populations of candidate nanoscale sources. This is particularly important for integration strategies that rely on screening individual emitters from bulk material or nanoparticles for assembling in photonic devices.[22–24]

Identifying genuine single-photon emitters generally requires measuring the second-order autocorrelation function, $g^{(2)}(\tau)$, of each potential candidate. This is done using a Hanbury Brown-Twiss (HBT) interferometer,[25,26] which measures the temporal correlations between photons detected at the two outputs of a beam splitter (Figure 1). The normalized value at zero delay, $g^{(2)}(0)$, quantifies the purity of the emission. Ideally, a single quantum emitter shows $g^{(2)}(0) = 0$, while an ensemble of $N$ independent, identically contributing emitters shows a value of $g^{(2)}(0) = 1 - 1/N$.[27] In practice, however, $g^{(2)}(0)$ is determined by fitting a parametric model to the histogram of measured coincidences, making the statistical uncertainty of the estimate inherently dependent on the number of detections. As the coincidence rate scales with the square of the single-photon emission rate, acquiring sufficient counting statistics can require long measurement times, especially for emitters and detection systems with low photon-extraction efficiency. This sets a fundamental trade-off between measurement throughput and number of practically surveyable candidate emitters.

Recent studies on quantum emitters have shown that machine learning (ML) can alleviate the compromise between data scarcity and measurement throughput,[28,29] including specifically for characterizing single photon sources.[30] Kudyshev et al.[30] showed that supervised ML classifiers—most notably a convolutional neural network (CNN)—can distinguish single-photon emitters from non-single emitters using sparse autocorrelation histograms. Using simulated and real nitrogen-vacancy (NV) center data, they show that their trained CNN reaches high classification accuracy from roughly a second of integration time, where conventional Levenberg–Marquardt (LM) fitting performs only slightly better than chance, and requires substantially longer acquisitions to reach comparable performance.

Here, we extend these results in two complementary directions. First, we introduce sequential Bayesian inference as a third classification approach alongside the established Levenberg–Marquardt and neural-network methods. Second, we use these three approaches to directly compare two competing inference paradigms: physics-based inference and data-driven inference.

The choice of a Bayesian approach is motivated by the way HBT autocorrelation data are acquired experimentally. Because the coincidence histogram accumulates incrementally throughout the measurement, each newly acquired set of coincidence counts can be treated as additional evidence that updates the posterior probability of each hypothesis, rather than requiring the entire dataset to be refitted at every integration time of interest. This naturally exploits the sequential structure of the measurement while retaining the physical interpretability of an explicit $g^{(2)}(\tau)$ model. Together with conventional LM fitting, the Bayesian approach constitutes the paradigm of physics-based inference, in which each classification is derived from explicit, physically interpretable model parameters. This physics-based paradigm allows us to make a direct comparison with data-driven inference, represented here by a feedforward neural-network classifier, whose use is motivated by previous demonstrations[30] that learned representations can outperform direct fitting under sparse-count conditions, albeit at the expense of an explicitly interpretable decision process.

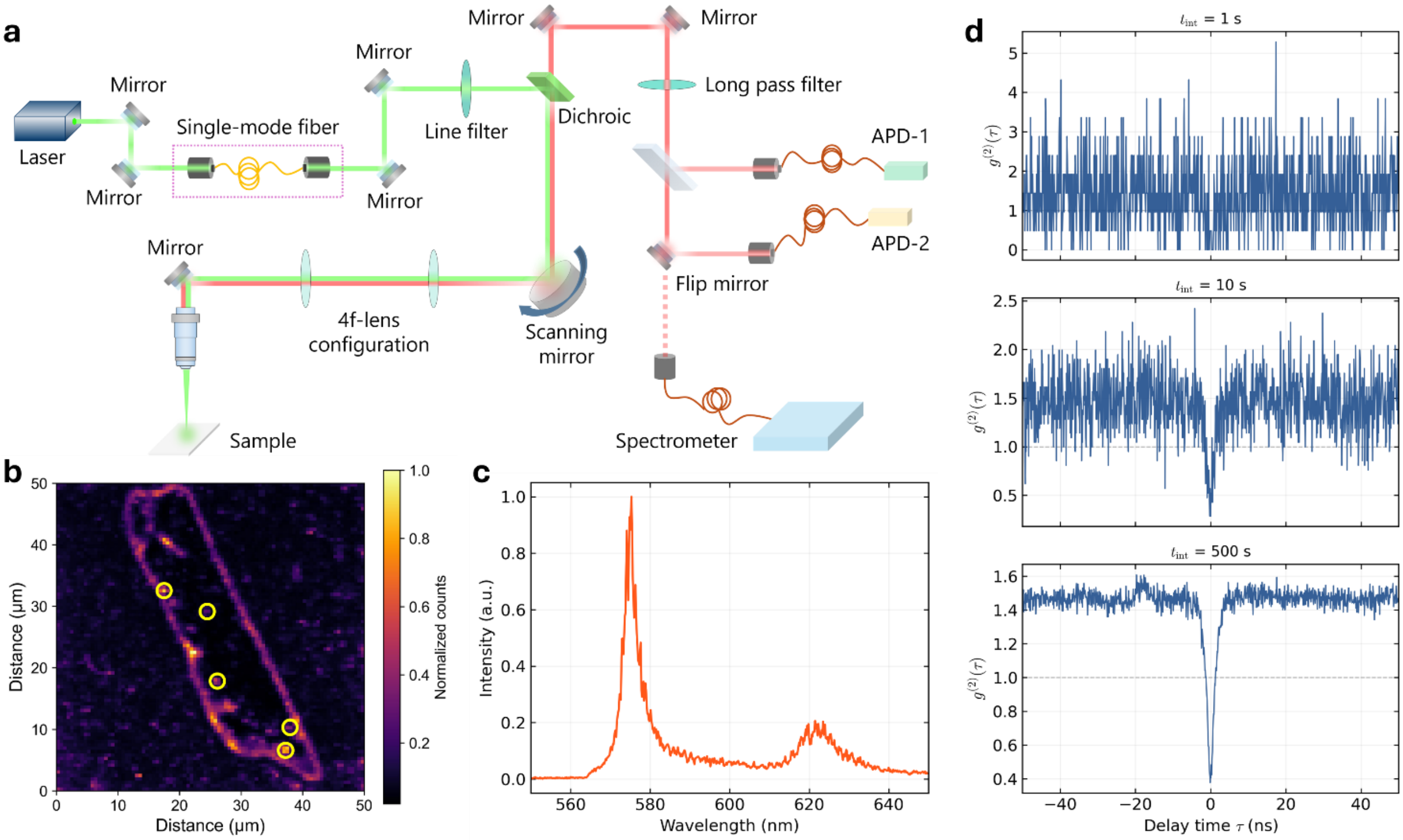

**Figure 1.** Experimental setup and measurements. **a)** Schematic of the experimental setup consisting of a confocal microscope, spectrometer, and Hanbury Brown–Twiss (HBT) interferometer. **b)** Confocal scan of the hBN flake showing bright emitters (circles). **c)** Photoluminescence (PL) spectrum of a representative emitter. **d)** Consecutive second-order autocorrelation measurements acquired with the experimental setup after integration times of 1, 10 and 500 s, illustrating the progressive accumulation of coincidence events.

To measure the performance of the three algorithms, we use synthetic training and test data calibrated using real experimental measurements. This enables the three approaches to be compared against an exactly known ground-truth emitter number while preserving realistic background levels, detection noise, and photon-extraction efficiencies.

We show that across the range of integration times considered, all three approaches ultimately achieve high, and often near-perfect, classification accuracy once sufficient data have been acquired, but differ substantially in how rapidly they approach this regime and in their robustness at the shortest integration times, where photon statistics are most limited. Specifically, LM is the most susceptible to failure under extreme sparsity and the slowest to reach consistently reliable classification. The Bayesian approach converges to high, stable accuracy fastest, once enough sequential evidence has accumulated, owing to its ability to combine information across integration time in a way single-shot refitting cannot. The neural network is markedly the most robust at the shortest integration times tested, avoiding the failure mode shared by the two physics-based approaches, though its classifications cannot be traced to an explicit physical quantity. Together, these results highlight practical trade-offs between classification performance, convergence speed, and interpretability, providing guidance for selecting an appropriate approach for different emitter-screening applications.

## 2. Materials and Methods

To benchmark the three classification approaches under realistic single-photon-source characterization conditions, we combine experimental Hanbury Brown–Twiss measurements with large-scale synthetic data generation. Key experimental parameters—most notably the background-to-signal ratio contributing to the measured coincidence rate—are extracted directly from real HBT measurements and used to calibrate a Monte Carlo emitter-and-detection model. This model generates large synthetic datasets where the number of emitters is known exactly, providing the ground truth required for robust cross-validated benchmarking of all three classifiers. This combined approach allows us train and evaluate every method on data reflecting realistic experimental noise, background, and photon extraction and collection inefficiencies, while preserving exact knowledge of the underlying emitter population (which is instead not known for real data) to rigorously assess classification performance. This approach addresses the fundamental challenge in benchmarking emitter-classification methods where experimental datasets provide realistic measurement statistics but lack a definitive ground-truth emitter count.

### *2.1. Experiment*

Experimental autocorrelation measurements were acquired from emitters in hexagonal boron-nitride (hBN). Silicon substrates with 285-nm thermally grown $SiO_2$ layers were sequentially cleaned in acetone and isopropyl alcohol (IPA) for 5 minutes each, followed by drying with an air gun. Then, bulk hBN crystals (purchased from HQ Graphene) were mechanically exfoliated onto the as-cleaned substrates using adhesive tape (Nitto SPV224). To activate quantum emitters, the samples were consequently annealed at 1000 ºC under an oxygen flow of 200 sccm. During annealing, the samples were placed at the center of the tube. The temperature was ramped at approximately 8 ºC/min, held at 1000 ºC for 1 hour, and then naturally cooled to room temperature.

Confocal maps, photoluminescence (PL) spectra, and second-order autocorrelation histograms were measured using a lab-built setup (Figure 1a). A continuous-wave green laser (Cobolt Samba, 532 nm) was employed as the excitation source. To maintain the spatial mode quality, the laser was coupled into a single-mode optical fiber before being focused on the sample surface through a 4f system and a 100× objective (NA = 0.7; MY100X-806, Thorlabs). In the excitation pathway, a laser-line filter (Semrock, LL01-532-12.5) was placed to ensure monochromatic 532 nm excitation. In the collection path, a long pass filter (Semrock, LP02−561RE-25) tilted by 30º was employed to reject any residual leakage of the 532-nm laser. During the measurement, a scanning mirror (Newport SFM-CD300B) steered the laser beam across the sample while an avalanche photodiode (APD1) detected the incoming photons to create a confocal map, and a spectrometer acquired the PL spectra (Figures 1b and 1c). Autocorrelation measurements were carried out employing a Hanbury Brown–Twiss interferometer, with different integration times (Figure 1d). A flip mirror was used to switch between the spectrometer and the HBT optical paths. All the photoluminescence (PL) spectra were collected at room temperature under an excitation power of 300 μW (measured at the back of the objective) with exposure time of 30 s. The $g^{(2)}(\tau)$ datasets were recorded at lower excitation power (100 μW) to reduce bunching. For each emitter, $g^{(2)}(\tau)$ traces were acquired with integration times following the 1-2-5 series: 1, 2, 5, 10, 20, 50, 100, 200 and 500 s.

*2.2. Synthetic data*

Real HBT measurements were acquired as coincidence histograms at a series of increasing integration times, each spanning a fixed delay window (±125 ns) and binned at fixed resolution (0.1 ns). Because the two detector channels in the HBT interferometer introduce a fixed timing offset (18.1 ns), all real histograms were first corrected by identifying and removing this offset and re-centering the antibunching dip at zero delay.

Both the real and the synthetic data are modelled with the reduced autocorrelation function,

$$g^{(2)}(\tau) = 1 - A_1 e^{-|t|/\tau_1} + A_2, \tag{1}$$

where $A_1$ and $\tau_1$ describe the antibunching dip amplitude and characteristic lifetime, and $A_2$ is a constant background pedestal. The model omits a bunching decay term, $e^{-|t|/\tau_2}$, since the accessible delay window is short compared to the bunching timescale of the emitters studied, making bunching indistinguishable from a constant offset within that window. However, the instrument's own long-timescale normalization is used to fix the $g^{(2)}(\tau)$ baseline to 1 in the experimental data rather than re-deriving it from the finite measurement window itself, avoiding a systematic bias that residual bunching would otherwise introduce.

For $N$ independent, equally contributing emitters diluted by an uncorrelated background at fraction $f_{\text{bg}}$ (signal purity $\rho = 1 - f_{\text{bg}}$), the zero-delay value follows $g^{(2)}(0) = 1 - \rho^2/N$.[27] Here, $f_{\text{bg}} = B/(S+B)$ is the complement of purity, where $S$ and $B$ are the signal and background contributions to the detected coincidence rate. Rather than measuring these directly, $f_{\text{bg}}$ is recovered by inverting this same relation. Given a dataset's known emitter number $N$ (established via a classical amplitude threshold on $g^{(2)}(0)$) and its measured $g^{(2)}(0)$, we determine $\rho = \sqrt{\max\left(0, N\left(1 - g^{(2)}(0)\right)\right)}$, which in turn gives the extracted background fraction, $f_{\text{bg}} = 1 - \rho$, specific to that measurement.

Synthetic training and test data are generated with a Gillespie-algorithm Monte Carlo simulation of a three-level emitter (ground, excited, and shelving states), producing simulated photon-detection timestamps that are split between two virtual detectors, mixed with an uncorrelated background click stream, and binned into coincidence histograms exactly as in the real measurement. Rather than assuming an idealized, noise-free background level, the background fraction injected into each simulated realization is determined from the distribution of $f_{\mathrm{bg}}$ values extracted from the real measurements above, so that the simulated data reproduces the same background statistics as the real experiment. Each simulated realization is assigned a ground-truth emitter number $N$, drawn to populate both the *single*-emitter ($N = 1$) and *multi*-emitter ($N \geq 2$) classes, and its histogram is accumulated over a sequence of increasing integration times, mirroring how real HBT data is acquired.

For training and evaluation, we focus on the binary discrimination between *single* ($N = 1$) and *multi* ($N \geq 2$) emitters, comparing each classifier's predicted class with the ground-truth class—derived from the known emitter number in the synthetic data. All three classifiers are evaluated using the same stratified *k*-fold cross-validation protocol. In each fold, the classifier is calibrated or trained exclusively on the training split and then evaluated on the held-out datasets at every integration time of interest. Consequently, every dataset is classified exactly once by a model that was neither calibrated nor trained using that file.

The Supplementary Information (SI) contains additional technical details on the handling of the experimental data (SI, Section S1.1), generation of the synthetic data (SI, Section S1.2), and implementation of all three methods (briefly summarized below), together with sensitivity analyses examining the effect of their respective design choices (SI, Sections S2–S4).

### *2.3. Levenberg–Marquardt approach*

The Levenberg–Marquardt (LM) approach directly fits the reduced autocorrelation model of equation (1) to each coincidence histogram via weighted nonlinear least-squares, with each bin weighted by its inverse Poisson variance to correctly account for the shot noise inherent to low-count data.[31] The fitted parameters yield an estimate of the zero-delay autocorrelation, $g^{(2)}(0) = 1 - A_1 + A_2$, which is compared against a decision threshold to classify the emitter as *single* or *multi*.

Note that the decision threshold is not fixed a priori but calibrated empirically from the training data at each fold: the fitted $g^{(2)}(0)$ values for known *single* and *multi* training datasets are used to determine the two classes' typical $g^{(2)}(0)$, and the threshold is set at the midpoint between them. Because fit behavior changes systematically with the amount of data available, a separate threshold is calibrated at each integration time considered, rather than calibrating once from the most-converged data and applying that single threshold throughout. To reduce iteration-to-iteration variability arising from independently refitting the model at each integration time, the classification margin (i.e. the fitted $g^{(2)}(0)$ relative to its corresponding threshold) is further smoothed across increasing integration time using an exponentially weighted moving average, before the final classification is made.

*2.4. Bayesian approach*

The Bayesian approach treats classification as sequential inference over the two hypotheses, *single* and *multi*, updated as new coincidence data accumulates. Because the raw coincidence histogram grows cumulatively over the course of a measurement, the counts newly added between two successive integration times are recovered by differencing the cumulative histograms at those two times, yielding a new, conditionally-independent increment of evidence at each step.

For each such increment, a reduced, single-parameter version of the model in equation (1) is used to obtain a per-increment estimate of $g^{(2)}(0)$: the antibunching lifetime $\tau_1$ is fixed at a representative value estimated once per fold from well-converged training data, and only the antibunching amplitude is estimated per increment, via closed-form weighted linear regression. This reduction avoids the instability of fitting all model parameters to the very small number of counts contained in a single increment. The distribution of these per-increment $g^{(2)}(0)$ estimates is modeled as Gaussian, with mean and variance calibrated separately for each class and for each position in the integration-time sequence, from training data—reflecting that the reliability of each increment's evidence varies systematically with integration time. Classification proceeds by Bayesian updating: starting from an uninformative prior, the posterior probability of each hypothesis is updated multiplicatively as each successive increment's evidence is incorporated, and the final classification is taken as the hypothesis with greater posterior probability at the integration time of interest.

*2.5. Feedforward neural-network approach*

The third approach uses a feedforward multilayer perceptron (MLP) to classify emitters directly from the shape of the coincidence histogram, without an explicit physical model. To limit the number of input features while preserving the discriminative shape of the antibunching dip, each autocorrelation histogram is reduced to a compact feature vector retaining full bin-by-bin resolution within a narrow window around the zero-delay bin, together with a small number of coarser, averaged bins summarizing the flat far-delay region of the histogram that establishes the background level.

The network consists of two hidden layers with a modest number of units, trained with L2 regularization and an internally-held-out validation split used for early stopping, to limit overfitting given the comparatively small number of independent emitters available for training relative to the dimensionality of the input. Input features are standardized, and the minority class is oversampled during training to counteract the natural class imbalance (*single*/*multi* – 20%/80%) arising from the emitter-number distribution used to generate the synthetic data. The network outputs a class probability for *multi*, which is compared against a fixed decision boundary to classify each file; as with the Bayesian approach, the classification is further stabilized by combining evidence across integration time, here via smoothing of the network's own output probability, rather than by any change to the decision boundary itself.

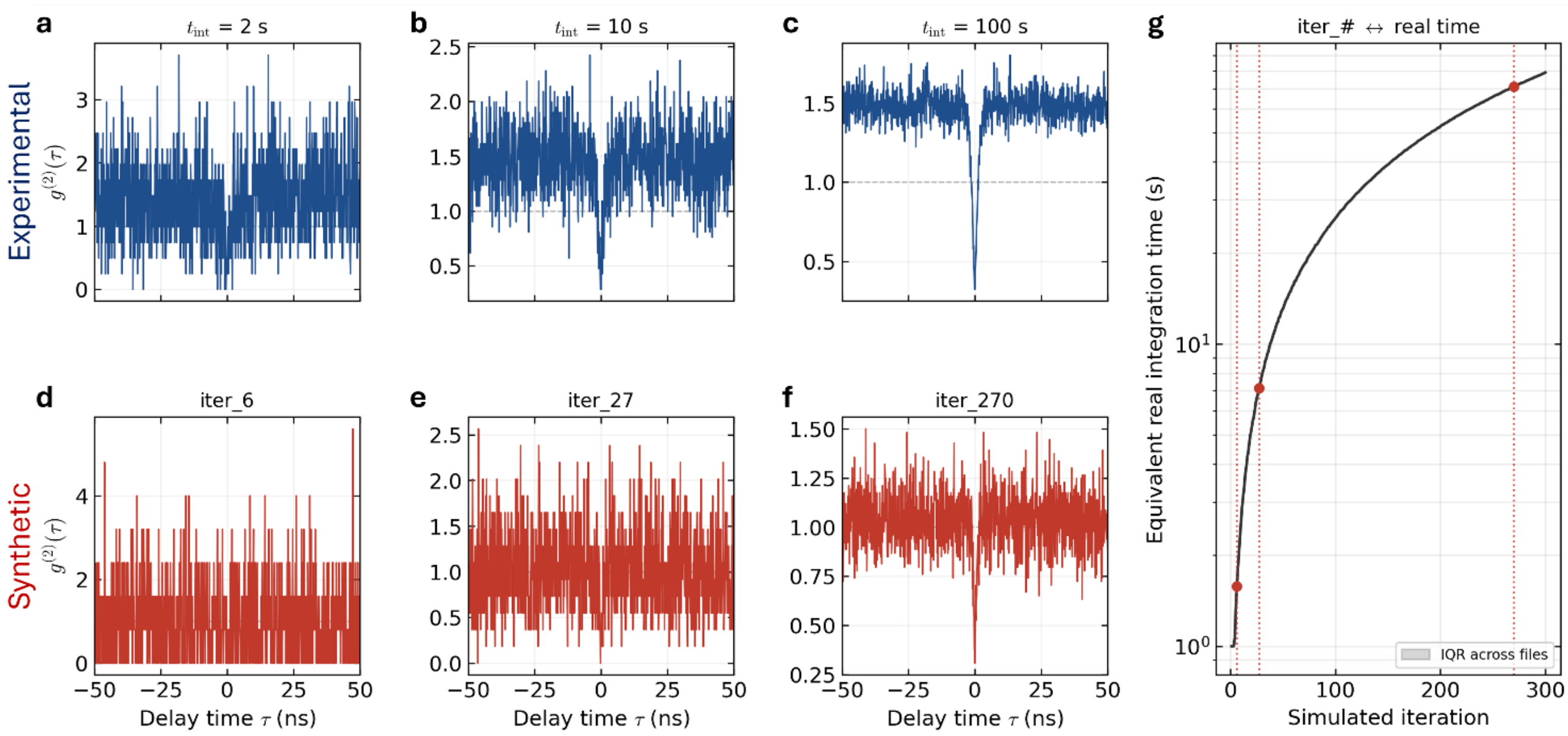

**Figure 2.** Correspondence between real experimental integration time and simulated iteration number. **a–c)** Experimental $g^{(2)}(\tau)$ autocorrelation traces from a single-emitter ($N = 1$) measurement, shown at three integration times (2, 10, and 100 s) spanning the sparse-to-converged range considered in this work. **d–f)** Synthetic $g^{(2)}(\tau)$ traces from a single-emitter chosen with an injected background fraction closely matching the real calibrated value, shown at the three simulated iterations whose baseline count rate corresponds to the real integration times in (a–c), via the mapping in (g). **g)** Correspondence between simulated iteration number and equivalent real integration time, obtained by matching the far-field baseline count rate between multiple real and simulated datasets (interquartile range across files: <2% of the mapped value at nearly all iterations, and not visually distinguishable from the curve at this scale); the vertical lines and markers indicate the three iterations shown in (d–f). All traces are normalized to $g^{(2)}(\tau)$; the dashed horizontal line marks the uncorrelated reference level $g^{(2)}(\tau) = 1$.

## 3. Results and Discussion

To benchmark the three approaches, we generated 5,000 synthetic autocorrelation datasets using the Monte Carlo simulation calibrated with experimentally derived noise-floor and photon-extraction parameters, as described in Section 2.2 (see also SI, Section S1). Of these, 1,002 correspond to *single* emitters, while the remaining 3,998 represent *multi* emitters (containing between two and five emitters). Each dataset consists of 300 cumulative autocorrelation histograms corresponding to iterations 1–300, where each successive histogram contains all coincidence events accumulated up to that iteration. This emulates experimental HBT measurements, in which the coincidence histogram builds up as the integration time increases. All three approaches are evaluated using the *k*-fold ($k$ = 5) stratified cross-validation scheme described in Section 2.2. The cross-validation split is performed at the dataset level; the held-out, test datasets are evaluated at 30 integration-time points (iterations 10–300 in steps of 10).

Figures 2a–c show experimental $g^{(2)}(\tau)$ traces measured at integration times 2, 10 and 100 s using our HBT experimental setup; figures 2d–f display the equivalent synthetic traces

based on the approximate mapping between simulation iteration and integration time estimated for our HBT setup and shown in Figure 2g. Note that at the shortest integration times, each histogram bin typically contains only a few coincidence events (often 0–2). Consequently, normalization by an equally sparse baseline yields only a small number of possible $g^{(2)}(\tau)$ values, giving the autocorrelation function a visibly discrete rather than continuous appearance (see also SI, Section S1.2, 'Thinning'). This is an expected consequence of finite counting statistics and is observed equally in both the experimental and simulated data.

### *3.1. Evaluation metrics*

To evaluate and compare the performance of the three approaches we test them on identical held-out (i.e. unseen) autocorrelation data using the same three metrics, *recall*, *precision* and *leakage,* with $N = 1$ (*single*) as the positive class, matching the practical goal of verifying genuine single-photon emitters:

$$recall = \frac{TP}{TP + FN} = P(\text{predict } single \mid \text{true } single), \tag{2}$$

$$precision = \frac{TP}{TP + FP} = P(\text{true } single \mid \text{predict } single), \tag{3}$$

$$leakage = \frac{FP}{FP + TN} = P(\text{predict } single \mid \text{true } multi), \tag{4}$$

where $TP$ is a true *single* correctly classified as *single*, $FN$ a true *single* misclassified as *multi*, $FP$ a true *multi* misclassified as *single*, and $TN$ a true *multi* correctly classified as *multi*.

Note that the *leakage* carries the standard single-photon-source-verification meaning of the term, i.e. the rate at which a genuinely multi-emitter source is falsely certified as a verified single-photon source. It should desirably be low, and it is—arguably—the most practically consequential of the three metrics, for it quantifies exactly the failure mode a downstream application cannot tolerate, e.g. selecting emitters for device engineering and integration.

### *3.2. Classification performance across integration time*

Table 1 summarizes the performance of each method at two critical points: the sparsest (iteration 10) and the most-converged (iteration 300) integration times tested.

**Table 1.** Performance at the sparsest and most-converged integration times.

| Method | Iteration | Recall | Precision | Leakage |
|---|---|---|---|---|
| **LM** | 10 | 0.8134 | 0.5028 | 0.2016 |
| **Bayesian** | 10 | 0.8224 | 0.5186 | 0.1913 |
| **MLP** | 10 | 0.6317 | 0.7177 | 0.0623 |
| **LM** | 300 | 0.9870 | 0.9851 | 0.0038 |
| **Bayesian** | 300 | 0.9750 | 1.0000 | 0.0000 |
| **MLP** | 300 | 0.9641 | 0.9979 | 0.0005 |

At full convergence (iteration 300), all three methods perform well, yet with salient and different behaviors. The Bayesian algorithm is the clear winner as it reaches perfect *precision* and zero *leakage*, the MLP comes very close (*precision* 0.9979, *leakage* 0.0005), whilst LM trails both on *precision* and *leakage* (0.9851 and 0.0038 respectively) despite having the *highest* recall of the three (0.9870).
At the sparsest integration time (iteration 10) however, the ranking changes substantially: the MLP's *leakage* (0.0623) is—favourably—more than threefold lower than LM's (0.2016) and Bayesian's (0.1913), but its *recall* is markedly lower (0.6317 vs. 0.8134 and 0.8224).

This last observation however requires some scrutiny, as the straight up values can lead to misleading conclusions. LM and Bayesian's higher *recall* at iteration 10 is not, per-se, evidence of better sparse-data performance. It is the signature of a specific failure mode. Both methods compute an explicit ratio or parametric fit over a narrow window of the $g^{(2)}(\tau)$ curve. At very low counts, near-empty bins (i.e. with no or low detected coincidences) in that window are common, regardless of the true emitter number. This biases the fitted or estimated $g^{(2)}(0)$ toward looking like a deep antibunching dip for both true *singles* and true *multis*. The result is a partial collapse toward predicting *single* almost independent of the actual input, which mechanically inflates *recall* (a collapsed classifier trivially catches every true *single*) while simultaneously inflating *leakage* (it also, for the same reason, falsely certifies many true *multis*). The evidence for this being a collapse rather than bona fide performance is exactly the simultaneous elevation of both *recall* and *leakage*. A classifier extracting genuine partial information from sparse data, as the MLP does, instead shows the more honest pattern of moderate *recall* accompanied by low *leakage*. These results and relevant observations highlight the need to report all three metrics—*recall*, *precision* and *leakage*—simultaneously for the analysis to be meaningful and robust. For instance, from a practical standpoint, *recall* should never be reported or interpreted in isolation as it could lead to misconstrued conclusions, especially in the sparse-data regime; for a high *recall* alongside an equally high *leakage* is not a favourable result.

Figure 3 shows a more complete picture of the algorithms' performance with *recall*, *precision* and *leakage* of the three methods directly compared as they evolve across increasing iterations.

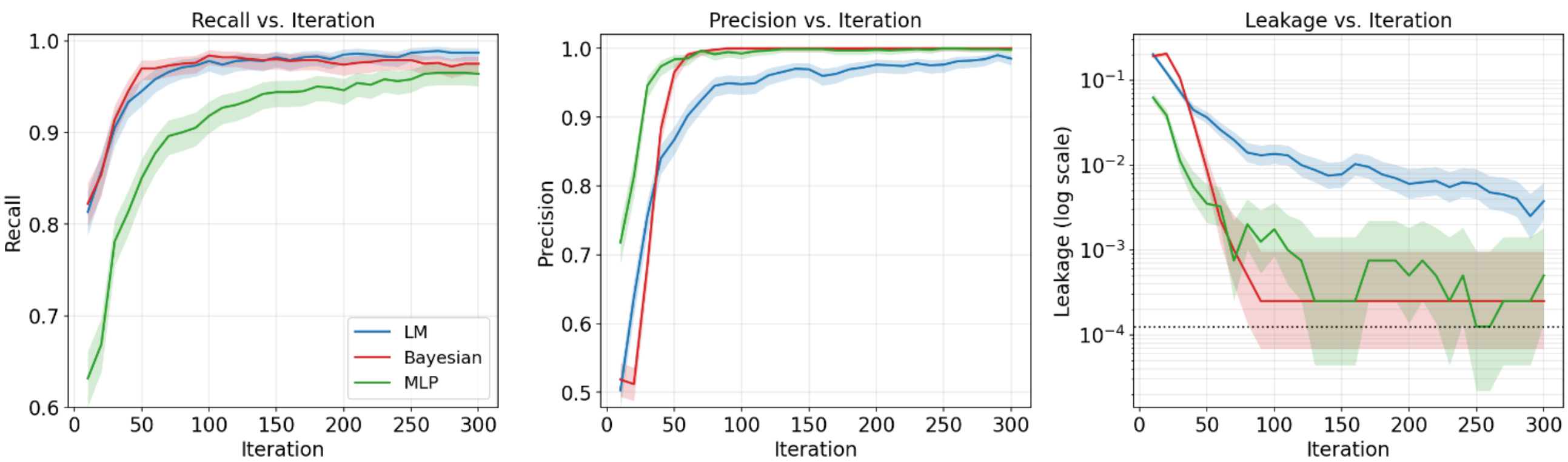

**Figure 3.** *Recall*, *precision*, and *leakage* as a function of integration time (iteration), for the Levenberg–Marquardt (LM), Bayesian, and MLP classifiers. *Recall* and *precision* axes are truncated to each panel's observed data range, rather than the full [0,1] scale, since no method's recall or precision falls below ~0.63 or ~0.50, respectively, at any integration time tested. *Leakage* is shown on a logarithmic scale as its convergence toward zero spans several orders of magnitude. Exact-zero leakage values are floored to half the smallest nonzero *leakage* value observed in the data, so they render at the bottom edge of the panel rather than being omitted; the black dotted line marks this floor. Shaded bands show the 95% Wilson score confidence interval at each iteration, computed directly from the held-out predictions; the Wilson interval is used in place of the standard normal approximation because it remains well-behaved when a method reaches exactly zero *leakage*, correctly reflecting that the true rate is only bounded, not proven to be exactly zero.

### *3.3. Convergence dynamics and stability*

Another behavior that warrants attention beyond the analysis of the two endpoints—i.e. the sparsest (iter. 10) and richest (iter. 300)—is the shape of each method as they approach convergence. This is summarized in Table 2 which indicates for each method two key figures of merit: the first iteration at which an algorithm reaches *precision* ≥0.99 and the number of *reversals* across iteration number. The number of *reversals* counts directional ‘zig-zags’ in the *precision*-vs-iteration curve, calculated between all 29 consecutive pairs of iterations. It is a measure of point-to-point jaggedness of *precision,* independent of its overall upward trend. Again, the Bayesian approach is a clear winner here as its *precision*-vs-iteration curve is nearly perfectly smooth with a single *reversal* across 29 steps. This is a structural consequence of sequential evidence chaining, and a further confirmation that the approach is inherently befitting the nature of HBT acquisitions. Specifically, in the Bayesian approach each posterior update multiplies in new, independent evidence on top of an accumulating log-likelihood: a process that cannot reverse sharply by construction. Conversely LM and MLP, which both re-derive an estimate independently at every iteration with no memory of previous iterations, are both exposed to per-iteration idiosyncratic noise that chaining is immune to. Correspondingly, both show substantially more jagged curves, with 9 and 13 total *reversals* for LM and MLP, respectively.

Bayesian's convergence profile also shows a characteristic ‘slow-then-sudden’ shape: *precision* remains comparable to LM's through the sparsest iterations (Table 1), then crosses 0.99 by iteration 60 and never leaves that regime again (Table 1, iteration-300 *leakage* of exactly zero). This is consistent with additive log-posterior accumulation: while individual increments carry weak evidence, progress is slow, but once evidence quality crosses a threshold, each additional increment's contribution compounds multiplicatively rather than merely averaging in as one more independent sample would for LM or MLP.

**Table 2.** Convergence speed and curve smoothness.

| Method | First iteration reaching *precision* ≥ 0.99 | Precision-curve reversals (out of 29 consecutive pairs) |
|---|---|---|
| **LM** | never reached (max. 0.99 at iter. 290) | 9 |
| **Bayesian** | 60 | 1 |
| **MLP** | 70 | 13 |

*3.4. Non-eliminable error floor*

Notably, none of the methods reaches perfect *recall* in the tested range, including at the most-converged integration time, i.e. at iteration 300. Converting the iteration-300 *recall* values from Table 1 back to raw counts: LM misses 13 of 1002 true *single*s (1.3%), Bayesian misses 25 (2.5%), and MLP misses 36 (3.6%). This is not a sampling artifact due to small evaluation set size: with 1002 held-out *single* datasets, the standard error on a recall of 0.97–0.99 is on the order of 0.005-0.01 (see Section 3.5), so these counts represent a real, quantifiable floor rather than noise from a handful of outliers.

To understand whether this floor reflects a shared set of legitimately ambiguous simulated emitters or independent, method-specific failures, we cross-referenced which specific datasets each method misclassified at iteration 300 (Table 3).

**Table 3.** Overlap in misclassified true-*single* datasets across methods at iteration 300.

| Method | Total missed | Unique to this method | Shared with ≥1 other method |
|---|---|---|---|
| **LM** | 13 | 1 (8%) | 12 (92%) |
| **Bayesian** | 25 | 10 (40%) | 15 (60%) |
| **MLP** | 36 | 15 (42%) | 21 (58%) |

Across all three methods: 4 datasets are missed by all three methods simultaneously, 18 are missed by two of three, and 26 are missed by one method alone (union of 48 distinct datasets across 74 total misses; see SI Section S5, Figure S1, for a visual representation). This suggests a simultaneous mixture of the two hypotheses, i.e., both legitimately ambiguous cases and method-specific blind spots. The 4 datasets missed universally are plausible candidates for ambiguous simulated draws, possibly due to parameter combinations that produce $g^{(2)}(\tau)$ curves none of the hereby tested classifiers could reliably resolve. However, a substantial fraction of each method's errors is not shared. Notably, Bayesian and MLP each have a distinct failure mode accounting for roughly 40% of their respective misses, while LM's failures are almost entirely (92%) failures shared with at least one other method (see Table 3), suggesting LM's error floor is closer to a subset of the genuinely ambiguous cases than a distinct failure of its own.

Because the three methods' misses are only partially correlated, combining their predictions is a natural next step. We thus construct a simple majority-vote ensemble that classifies a dataset as *multi* whenever at least two of the three methods agree. Under this rule, a true *single* is missed by the ensemble if and only if two or more of the individual methods miss it. From the vote-count breakdown underlying Table 3, this corresponds exactly to the 18 datasets missed by two out of three methods plus the 4 files missed by all three (18 + 4 = 22), while the 26 files missed by only a single method are correctly rescued by the other two. This reduces the total missed *singles* at iteration 300 from Bayesian's 25 or MLP's 36 down to 22, while matching MLP's low leakage (0.0005) rather than LM's higher leakage (0.0038)—a better-balanced result than any individual method achieves on its own, and a direct, quantified consequence of Table 3's overlap structure.

**Table 4.** Majority-vote ensemble performance vs. individual methods, iteration 300.

| Method | Recall | Precision | Leakage | Singles missed |
|---|---|---|---|---|
| **LM** | 0.9870 | 0.9851 | 0.0038 | 13 |
| **Bayesian** | 0.9750 | 1.0000 | 0.0000 | 25 |
| **MLP** | 0.9641 | 0.9979 | 0.0005 | 36 |
| **Majority vote** | **0.9780** | **0.9980** | **0.0005** | **22** |

### *3.5. Statistical robustness*

Several of the comparisons made in sections 3.1–3.4 are large and unambiguous. The sparse-regime leakage gap between MLP and LM (0.0623 vs. 0.2016) corresponds to a difference of roughly 18.8 standard errors given the sample sizes involved, ruling out chance. Other comparisons, however, are close enough that sample size matters. At iteration 300, LM's *recall* (0.9870) exceeds MLP's (0.9641) by approximately 3.3 standard errors, a reasonably solid difference, but LM vs. Bayesian (1.97 standard errors) and Bayesian vs. MLP (1.43 standard errors) are more marginal, lowering the confidence in the ranking between those two pairs. This is because *recall* is evaluated only over the true-*single* datasets, so its statistical precision is set by the number of *singles* available—1002 of the 5000 total files, *single* being the minority class in this ~20%/80% split. Resolving the LM-vs-Bayesian and Bayesian-vs-MLP rankings with confidence comparable to the other results reported here would require substantially more *single* datasets beyond the 1002 available.

*3.6. Strengths, weaknesses and trade-offs*

Table 5 summarizes in a qualitative manner the trade-offs we identified for each method as a result of the quantitative analysis we performed.

**Table 5.** Strengths, weaknesses, and best-fit use cases.

| Figure of Merit | LM | Bayesian | MLP |
|---|---|---|---|
| **Interpretability** | **Full** - every classification traces to explicit fit parameters | **Full** - explicit physical model, chained rather than single-shot | **None** - the model's decision cannot be attributed to a physical quantity |
| **Requires training** | **No** | **No** (only a one-off $\tau_1$ and likelihood calibration) | **Yes** - full supervised training pipeline |
| **Sparse-regime leakage** | **Poor** - most vulnerable to the collapse described in Section 3.2 | **Poor** - same collapse mechanism described in Section 3.2 | **Best** of the three, by a wide margin |
| **Convergence speed to high precision** | **Slowest** - never reaches 0.99 | **Fastest** - reaches 0.99 by iteration 60 | **Moderate** - reaches 0.99 by iteration 70 |
| **Curve smoothness** | **Moderate** – 9 reversals/29 pairs | **Best** - smooth by construction (1 reversal/29 pairs) | **Worst** - 13 reversals/29 pairs |
| **Recall at full convergence** | **Highest** (0.987) | **Moderate** (0.975) | **Lowest** (0.964) |
| **Computational cost** | **Highest** per fold (per-iteration threshold recalibration multiplies fitting cost) | **Low** per step (closed-form update) | **Moderate** - training cost, but fast inference once trained |
| **Distinct (method-specific) error mode** | **Minimal** (8% of its errors are unique to it) | **Moderate** (40%) | **Largest** (42%) |

Analysis of Table 5 leads to several key observations, the most apparent of which is that no single method or inference paradigm—physics-based or data-driven—dominates across all performance metrics. This leads to two broader conclusions. First, evaluating model performance using a single metric, such as *recall*, *precision*, or *leakage*, can be overly reductive and potentially misleading (see Section 3.2). Second, a hybrid strategy that combines multiple methods may provide the most effective overall approach by leveraging their complementary strengths while mitigating their individual weaknesses.

The preferred method therefore depends on the operational priorities of the application. If the primary objective is to minimize false verification at the shortest possible integration time—arguably one of the most important criteria for scalable single-photon-source screening—the MLP is the preferred choice, particularly at the sparsest integration times. If instead the priority is rapid and reliable convergence while retaining full physical interpretability, the sequential Bayesian approach is preferable, reaching near-perfect precision faster than either alternative while remaining grounded in an explicit and auditable $g^{(2)}(\tau)$ model. LM remains valuable as a no-training, fully transparent baseline, and its performance demonstrates that the simplest and most established approach is not necessarily the least capable. Indeed, once sufficient data have accumulated, LM achieves the highest *recall* of the three methods, albeit with slower convergence and greater *leakage* at sparse integration times.
Therefore, where computational resources permit, combining predictions across methods, even through a simple majority-vote scheme, could capture a meaningful portion of their complementary strengths, at the practical cost of running all three classifiers.

## 4. Conclusions

In this work, we benchmarked three complementary classification strategies—Levenberg–Marquardt fitting, a newly-proposed sequential Bayesian inference approach, and a feedforward neural network—for distinguishing single- from multi-photon emitters from second-order autocorrelation data. To assess performance, we used synthetic datasets calibrated against real HBT measurements from hexagonal boron nitride quantum emitters, enabling rigorous, ground-truth-aware benchmarking under realistic experimental noise, background, and photon-extraction conditions.

We showed that all three approaches ultimately achieve high, often near-perfect, classification accuracy once sufficient integration time has elapsed, but differ substantially in how they reach this regime.
The neural network is the most robust classifier at the shortest integration times tested, avoiding a partial collapse toward predicting *single* under extreme sparsity that, conversely, affects both physics-based approaches. This makes it the preferred choice when minimizing false verification at short integration times is the priority.
The sequential Bayesian classifier combines the fastest, smoothest convergence to near-perfect accuracy with full physical interpretability. This is a direct consequence of its ability to chain genuinely independent increments of evidence as the measurement accumulates; a property intrinsic to how HBT data is acquired experimentally and not shared by either the neural network or LM fitting.
LM remains a valuable, no-training, fully interpretable baseline: although the slowest to reach reliable classification and the most vulnerable to the sparse-count failure mode, it achieves the highest *recall* of the three methods once sufficient data have accumulated.

None of the three methods eliminates classification error entirely: a small, quantifiable fraction of single emitters (1.3-3.6%, depending on the method) remains misclassified even at the longest integration time tested. Cross-referencing which specific datasets each method misclassifies shows these failures to be only partially correlated, suggesting a mixture of legitimately ambiguous cases and method-specific blind spots. A simple majority-vote combination of all three classifiers exploits this partial independence, reducing the total number of misclassified single emitters beyond what any individual method achieves alone.

More broadly, our results show that no single method or inference paradigm—physics-based or data-driven—dominates across every performance axis considered, and that evaluating a classifier by a single metric such as *recall* in isolation can obscure a real underlying trade-off.

We anticipate that the calibrated synthetic-data benchmarking approach used here, together with the qualitative trade-offs identified between interpretability, convergence speed, and sparse-data robustness, will provide practical guidance for selecting or combining classification strategies for scalable single-photon-emitter screening across other solid-state emitter platforms.

## 5. Supporting Information

Experimental data handling, correction and parameters extraction for data simulation (Section S1.1). Monte Carlo simulation and synthetic data generation (Section S1.2). Levenberg–Marquardt model implementation (Section S2). Sequential Bayesian model implementation (Section S3). Feedforward neural-network (MLP) model implementation (Section S4). Supplementary figures, overlap in misclassified true-*single* datasets across the three models (Section S5).

## 6. Contributions and Acknowledgements

Collection, curation and analysis of the data was performed by N.M.N., M.S.H., D.A.N. and T.T.T. Theoretical modelling, and analysis and interpretability of the data rationale were provided by C.B. All authors contributed to writing, editing and reviewing the manuscript.

The Natural Sciences and Engineering Research Council of Canada (DGECR-2021-00234 and NSERC DISCOVRY 3/26) and the Canada Foundation for Innovation (John R. Evans Leaders Fund #41173) are acknowledged. T. T. T acknowledges the financial support from the Australian Research Council (DE220100487). C.C acknowledges the financial support from the Australian Research Council (DE250100406). This research is supported by an Australian Government Research Training Program (RTP) Scholarship.

## 7. Conflict of interest

The authors declare no conflict of interest.

## 8. Data Availability

The data and source code supporting the findings of this study are available in the public repository linked below. The repository includes: (i) the preprocessing pipeline used to analyse the experimental HBT measurements (also included) and extract the experimental parameters required to calibrate the synthetic data generation; (ii) the Monte Carlo simulation used to generate the synthetic autocorrelation datasets; and (iii) the implementations of the three classification approaches evaluated in this work—Levenberg–Marquardt (LM), sequential Bayesian inference, and the multilayer perceptron (MLP)—to classify the number of emitters in each synthetic dataset.

https://github.com/QML-trentu/Autocorrelation

# Supplementary Information

## Physics-Based versus Data-Driven Classification of Single-Photon Quantum Emitters from Sparse Autocorrelation Data

*Nhat Minh Nguyen,[1] Md Shakhawath Hossain,[1] Duc Anh Ngo,[1] Chaohao Chen,[1] Xiaoxue Xu,[1] Toan Trong Tran[1,*] and Carlo Bradac[2,*]*

[1] School of Electrical, Mechanical and Biomedical Engineering, University of Technology Sydney, Ultimo, NSW, 2007, Australia

[2] Department of Physics & Astronomy, Trent University, 1600 West Bank Dr., Peterborough, Ontario K9L 0G2, Canada

*Corresponding author, e-mail: carlobradac@trentu.ca
*Corresponding author, e-mail: trongtoan.tran@uts.edu.au

## S1. Synthetic data generation

### S1.1. Correction and background extraction from real data

Real HBT measurements are acquired as cumulative coincidence histograms at nine integration times following a 1-2-5 progression: 1, 2, 5, 10, 20, 50, 100, 200, and 500 s. Each raw file additionally records the single-channel count rate on each detector (APD1, APD2) at every integration time, used below to establish a reliable normalization baseline.

**Timing correction.** The two detector channels introduce a small, fixed timing offset between them, so the antibunching dip does not occur at $\tau = 0$ in the raw data. For each file, this offset is detected from the longest-integration (most-converged) histogram: the histogram is tail-normalized (mean of $|\tau| \geq$ 10 ns), smoothed with a 15-bin centered rolling average, and the offset is taken as the $\tau$ location of the minimum smoothed value within a 2-50 ns search window (chosen to safely exclude $\tau = 0$ itself while still capturing the expected instrument offset range). The same shift is then applied uniformly to every integration-time column of that file, since the offset is a fixed hardware property rather than something that varies with acquisition time.

**Baseline normalization.** Two sources of baseline are used, in order of preference: (1) if a separately-normalized companion file from the acquisition software (Time Tagger, Swabian Instruments) is available, the ratio of raw to normalized counts gives the instrument's own applied baseline directly (confirmed constant to six decimal places across all bins within a file); (2) otherwise, the baseline, $b$, is computed from the accidental-coincidence formula, $b = r_{\text{APD1}} \cdot r_{\text{APD2}} \cdot \Delta\tau \cdot T$, which reproduces the instrument's own normalization to within 1-3% where both are available. In this formula, $r_{\text{APD1}}$ and $r_{\text{APD2}}$ are the detection rates of the two photodetectors, $\Delta\tau$ is the bin width and $T$ is the integration time. This baseline is preferred over deriving a baseline from a narrow window of the measurement itself, since the latter would be biased downward by residual bunching that has not yet decayed within the ±125 ns raw acquisition window.

**Labeling.** Each file's true emitter number $N$ is derived from the timing-corrected, longest-integration data, using a classical amplitude threshold on the fitted or directly-estimated $g^{(2)}(0)$ (ideal floor $1 - 1/N$). Files whose classification-relevant $g^{(2)}(0)$ value falls within a small margin of a threshold boundary are excluded as genuinely ambiguous, rather than forced into a class.

**Background fraction extraction.** For each surviving, corrected dataset with known $N$, the fitted or directly-estimated $g^{(2)}(0)$ is inverted via $\rho = \sqrt{\max\left(0, N\left(1 - g^{(2)}(0)\right)\right)}$, with $f_{\text{bg}} = 1 - \rho$ (see main text Section 2.2), giving one extracted $f_{\text{bg}}$ value per real dataset. Files with $g^{(2)}(0) > 1$ are excluded from this step: since the underlying dilution relation is only valid for $g^{(2)}(0) \leq 1$, values above 1 would otherwise be silently clamped to a degenerate $f_{\text{bg}} = 1.0$ (100% background), which does not reflect a real measurement. Extracted $f_{\text{bg}}$ values are pooled by emitter number: $N = 1$ on its own (for the *single* class), and $N = 2$ through $5$ pooled together (for the *multi* class).

### S1.2. Monte Carlo simulation

Synthetic data is generated by a Gillespie-algorithm simulation of a three-level emitter (ground, excited, and metastable shelving states), independently for each of the 5000 realizations.

**Per-realization setup.** The number of emitters $N$ is drawn uniformly from 1 to 5 (giving an approximately 20%/80% *single*/*multi* split in the resulting dataset). Each individual emitter is independently assigned an antibunching lifetime $\tau_1$ drawn uniformly in the range 0.5–1.5 ns and a bunching decay time $\tau_2$ drawn uniformly in the range 2.0–6.0 ns; the shelving-state transition rate and the excitation rate are fixed constants with values 0.02 and 0.1, respectively.

**Photon generation and detection.** Photon arrival timestamps are generated chunk by chunk across 300 iterations, assuming 100 photons per emitter per chunk, and accumulated into a running coincidence histogram at each step—mirroring how a real integration-time series accumulates. Note that the choice of 100 photons is arbitrary; however, we conduct a full analysis to establish the correspondence between the iteration number and the real integration time (see main text, Figure 2g, Section 2.5). Within each chunk, generated photons are split 50/50 between two virtual detectors. Each photon's detection time is jittered by a Gaussian with standard deviation jitter (modeling detector timing jitter), and an uncorrelated background click stream is added to each detector independently at a rate, $r_{\text{B}}$, computed from that chunk's actual signal rate, $r_{\text{S}}$, and the realization's target background fraction, $f_{\text{bg}}$, as: $r_{\text{B}} = \left(f_{\text{bg}}/\left(1 - f_{\text{bg}}\right)\right) \cdot (r_{\text{S}}/2)$, where the factor of 2 correctly accounts for background being added per-detector after the 50/50 split, while the signal rate, $r_{\text{S}}$, is computed from the combined (pre-split) photon stream.

**Thinning.** The naturally-generated coincidence counts, summed across all 300 chunks, land orders of magnitude above what a real acquisition produces at any integration time. Before saving, each realization's full sequence of accumulated histograms is thinned via binomial thinning (which exactly preserves a Poisson distribution at a reduced mean, correctly adding the additional shot noise a lower-count real acquisition would have) to a target baseline of 60 counts/bin in the far-tail region at the final (most-accumulated) iteration. The thinning probability is computed once from the final iteration and applied identically to every earlier iteration snapshot, so the sparse-to-clean progression across iterations remains internally consistent. One consequence, addressed directly in the main text (main text, end of Section 3), is that at the very sparsest iterations, individual bins typically contain only 0-2 raw counts. Thus, normalizing by an equally sparse baseline yields a visibly discrete rather than continuous $g^{(2)}(\tau)$, which is an expected consequence of finite counting statistics, observed equally in experimental and synthetic data at comparably sparse integration times (see main text, Figure 2a, d).

**Self-consistency check.** After thinning, each realization's own final-iteration histogram is independently re-fit (5-parameter model, including offset and $\tau_2$, since the simulated bunching decay time is short and resolvable, unlike the real-data case; see main text, Section 2.2), to re-extract an effective $f_{\mathrm{bg}}$ and $g^{(2)}(\tau)$, which are saved alongside the injected ground-truth values for auditing.

## S2. Levenberg-Marquardt implementation

**Fitting.** At each evaluated iteration, the accumulated histogram is fit to the reduced model $g^{(2)}(\tau) = 1 - A_1 e^{-|t|/\tau_1} + A_2$ (main text, Eq. 1) via weighted nonlinear least-squares, with each bin weighted by its inverse Poisson variance ($\sigma = \sqrt{C/b}$, where $C$ and $b$ are the raw counts and baseline, respectively), derived from a companion raw-counts file saved alongside the normalized data) to correctly account for shot noise at low counts. Initial parameter guesses use percentile-based ($2^{nd}/90^{th}$) rather than raw min/max statistics, since raw extrema are easily thrown off by a single noisy bin at low counts, producing guesses outside the fit's own bounds and causing the optimizer to reject the fit.

**Cross-validation.** All three classifiers share an identical evaluation protocol: 5-fold stratified cross-validation, evaluated at 30 integration-time points (iterations 10, 20, 30, ..., 300). In each fold, only that fold's training files are used for any calibration or fitting parameter estimated from data; the held-out test files are evaluated using only those fold-specific, pre-calibrated quantities, so every file is classified exactly once by a model that never saw it during calibration.

**Threshold calibration.** The classification threshold is calibrated separately at each evaluated iteration (not once from the best-converged data and reused throughout), using that iteration's fitted $g^{(2)}(0)$ values for the fold's training files, grouped by true label: the threshold is the midpoint between the two classes' mean fitted $g^{(2)}(0)$ at that iteration. If a specific (class, iteration) combination has fewer than 5 successfully-fit training files, that iteration falls back to the pooled (all-iteration) mean for that class.

**Smoothing.** Predictions are further stabilized via exponential moving average (EMA, α = 0.4) smoothing of the classification margin—the fitted $g^{(2)}(0)$ relative to that iteration's calibrated threshold—rather than the raw fitted $g^{(2)}(0)$ value directly. The calibrated threshold itself shifts substantially across the iteration range (from ≈0.23 at the sparsest evaluated iteration to ≈0.49-0.50 at the most converged, in the dataset used for the main text), so smoothing raw $g^{(2)}(0)$ would blend evidence computed under very different decision boundaries before comparing it to only the current iteration's threshold. Smoothing the threshold-relative margin instead keeps this comparison internally consistent at every step.

**Sensitivity to these design choices.** Both refinements were identified and validated on smaller-scale test batches during development, then confirmed on the full 5000-file dataset used for the main text results. Per-iteration threshold calibration, versus a single threshold calibrated once from the best-converged iteration and reused throughout, substantially improved sparse-regime (iterations 10-90) precision and leakage in this earlier development-stage testing (see also the discussion of a specific sparse-regime collapse mechanism in the main text Section 3.2, which per-iteration calibration mitigates without eliminating). Margin-based EMA smoothing, versus the naive alternative of smoothing the raw fitted $g^{(2)}(0)$ value directly (the direct analog of the smoothing scheme used for the MLP, Section S4), was tested and rejected during the same development stage: raw-value smoothing improved *recall*'s curve smoothness but worsened *precision*'s—a genuinely mixed result, traced to the threshold-shift issue described above. Margin-based smoothing was adopted instead specifically because it resolved this without the corresponding downside; on the final 5000-file dataset, margin smoothing reduces LM's *precision*-curve *reversals* from 15 (unsmoothed) to 9 (Table 2), with final-iteration (iteration 300) *precision* improving from 0.9425 to 0.9851 and *leakage* from 0.0150 to 0.0038.

## S3. Sequential Bayesian implementation

**Motivation for chaining on increments, not cumulative values.** Because the simulated datasets iteration columns are cumulative (iteration $t$ contains all counts accumulated up to and including chunk $t$), a Bayesian posterior chained directly on these cumulative $g^{(2)}(\tau)$ estimates would double-count already-observed data at every update step, manufacturing artificial confidence. Instead, the counts added between two successive evaluated iterations are recovered by differencing the cumulative raw histograms at those two points, yielding a new, conditionally-independent incremental measurement at each of the 30 evaluated steps.

**Aggregation.** A single such increment (one 300$^{th}$ of the total accumulated data) is too sparse for a stable per-step estimate. Each of the 10 raw chunks between two consecutive evaluated iterations is summed into one "super-chunk" before extracting a per-step $g^{(2)}(0)$ estimate, so the same 30 evaluated iterations used by the other two methods are retained while ensuring each individual update carries a workable amount of evidence.

**Per-super-chunk estimator.** A full nonlinear fit at each super-chunk is unstable given how sparse even the aggregated per-step data is. In initial testing, many super-chunk fits pinned at the same parameter bound regardless of true class, leaving the two classes' likelihoods nearly indistinguishable. Instead, a closed-form, reduced estimator is used: the antibunching lifetime $\tau_1$ is fixed at a single representative value, estimated once per fold from a full fit to well-converged (last-iteration) training data, and only the antibunching amplitude is estimated per super-chunk via closed-form weighted linear regression against the fixed-$\tau_1$ model. This removes the extra degrees of freedom responsible for the instability while still using the data's actual shape, rather than, for instance, a simple windowed count ratio.

**Likelihood calibration.** The distribution of per-super-chunk $g^{(2)}(0)$ estimates is modeled as Gaussian, with mean and variance calibrated separately for each class and for each of the 30 chain positions (rather than pooled across the whole chain into one Gaussian per class), reflecting that the reliability of a super-chunk's evidence changes systematically over the course of the chain—an early, sparse super-chunk and a late, well-converged one are not equally-informative draws from the same underlying distribution. If a specific (class, chain-position) combination has fewer than 10 successfully-estimated training super-chunks, that position falls back to the pooled (all-position) estimate for that class.

**Sequential update.** Starting from an uninformative prior, the posterior probability of each hypothesis (*single*, *multi*) is updated multiplicatively (equivalently, additively in log-space) as each successive super-chunk's evidence is incorporated, using the per-position calibrated likelihood. Super-chunks flagged unreliable (insufficient baseline counts, or a non-convergent per-chunk fit) are skipped, carrying the posterior forward unchanged for that step. The final classification at a given evaluated iteration is the hypothesis with greater posterior probability at that point in the chain.

**Sensitivity to this design choice.** Per-chain-position likelihood calibration, versus a single pooled ($\mu$, $\sigma$) per class shared across the whole chain, was the single most impactful design change tested across all three classifiers in this work. On an earlier development test set, the pooled version needed a long 'warm-up' period before its *precision* exceeded LM's, only reaching *precision* 1.000 by iteration 200, while *leakage* at the sparsest tested iteration remained at 0.989 (effectively a total collapse) throughout that warm-up; per-position calibration removed this warm-up almost entirely, reaching *precision* 1.000 by iteration 70 and reducing *leakage* at the same sparsest iteration to 0.283. This structural property—fast, stable convergence once past a brief warm-up, and, uniquely among the three methods, a strictly monotonic-looking *precision* curve (1 *reversal* out of 29 consecutive iteration pairs in the final 5000-file results, Table 2)—is a direct consequence of sequential evidence chaining and is further discussed in the main text, Section 3.3.

## S4. Feedforward neural-network (MLP) implementation

**Feature representation.** Each accumulated histogram (2001 raw $\tau$-bins over the ±100 ns window) is reduced to a compact feature vector before being passed to the network: full original bin-by-bin resolution within ±10 ns of $\tau = 0$ (where essentially all the discriminating antibunching-dip shape lives), plus 15 coarse-averaged bins summarizing the flat far-delay region on each side (which only needs to establish the background level, not fine structure). This keeps the feature count small (~230, versus 2001 for the full histogram) relative to the number of independent datasets available for training, limiting the network's capacity to memorize dataset-specific noise rather than learn the general sparse-to-clean pattern. A uniform downsampling of the full histogram to a comparably small number of bins was tested and found to perform worse, since the resulting bin width becomes comparable to or wider than the narrowest $\tau_1$ in the simulated range, washing out the dip almost entirely—the same coarse-binning bias discussed for the LM fit in the main text, Section 2.2.

**Architecture and training.** A scikit-learn MLPClassifier with two hidden layers of 64 and 32 units, L2 regularization (alpha = $1\times10^{-2}$), and early stopping is trained independently for each fold, using only that fold's training datasets. Input features are standardized, and the minority (*single*) class is oversampled/duplicated with a small amount of Gaussian jitter per duplicate to balance classes before training, since the default ~20%/80% class split of the synthetic data otherwise lets the network achieve deceptively high overall training accuracy while performing poorly on the minority class. Note that oversampling is applied to training data only, never to the held-out test files, which are always evaluated at their true, natural class balance.

**Output and smoothing.** The network outputs a predicted probability of *multi*, compared against a fixed 0.5 decision boundary for classification. As with the Bayesian classifier, predictions are further stabilized by combining evidence across integration time—here via EMA smoothing ($\alpha$ = 0.4) of the network's own output probability sequence for each file—rather than any change to the decision boundary itself. This is explicitly not a Bayesian chain and makes no independence claim: the network is given the cumulative histogram directly at every iteration, so consecutive predictions are not independent evidence, and combining them the way the Bayesian classifier's per-super-chunk increments are combined would manufacture false confidence from heavily overlapping input data. EMA smoothing instead rests on a different, more modest justification: a dataset's true label does not change across iterations, so each iteration's prediction is a noisy read-out of a persistent, unchanging truth, and averaging multiple such read-outs is a standard variance-reduction technique.

**Sensitivity to these design choices.** On the 5000-file dataset used for the main text, EMA smoothing improved final-iteration (iteration 300) *precision* from 0.9928 to 0.9979 and *leakage* from 0.0018 to 0.0005, and mean *precision* across all 30 evaluated iterations from 0.9621 to 0.9786, though—unlike for LM—it did not reduce curve jaggedness (13 *reversals* both before and after smoothing, main text, Table 2), consistent with the MLP's decision boundary being fixed throughout rather than iteration-dependent as LM's calibrated threshold is.

## S5. Supplementary figures

**Figure S1.** Three-way overlap in misclassified true-*single* datasets at iteration 300, corresponding to the counts and percentages reported in Table 3 of the main text. Circle areas are not drawn strictly proportional to set size. LM: 13 total misses (1 unique to LM, 12 shared with ≥1 other method); Bayesian: 25 total misses (10 unique, 15 shared); MLP: 36 total misses (15 unique, 21 shared); 4 files are missed by all three methods simultaneously.

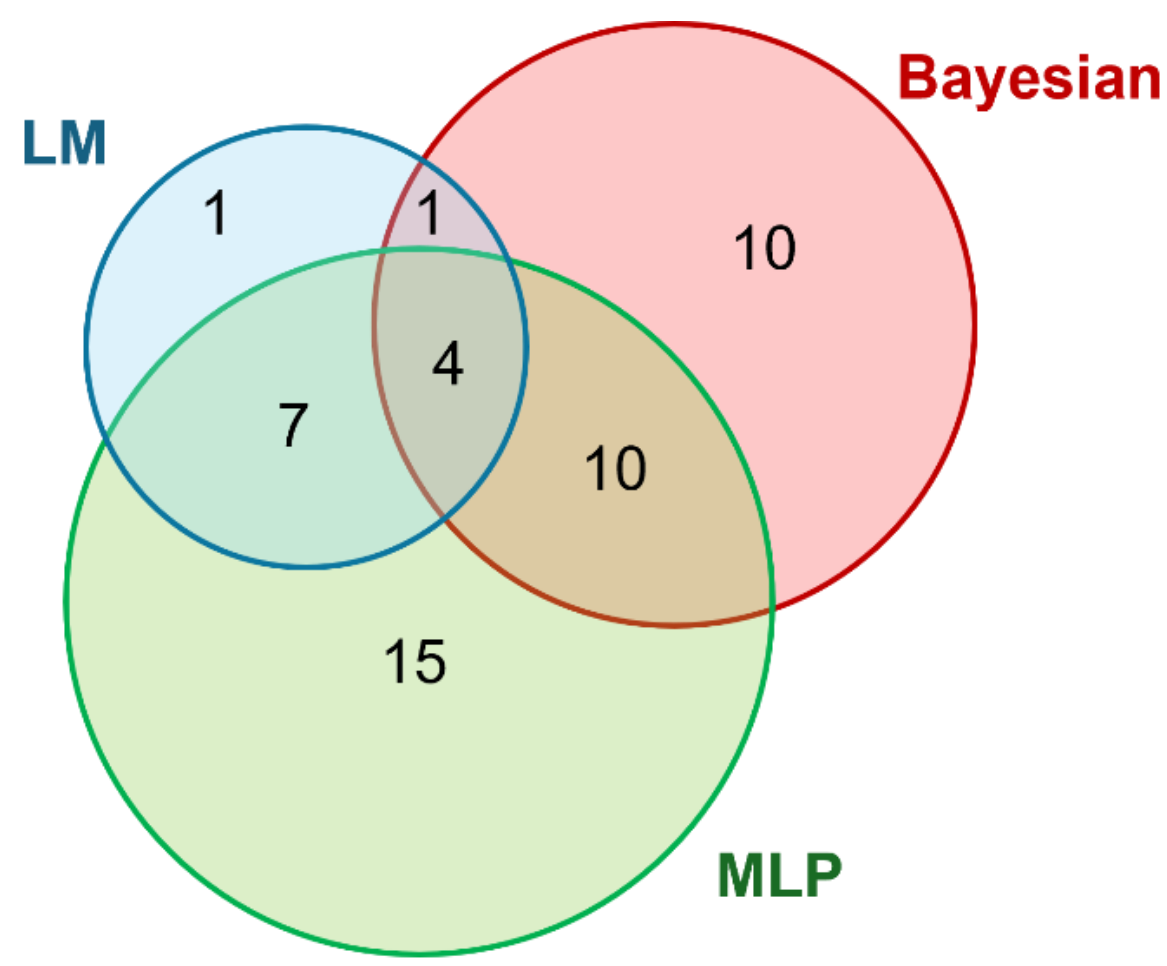

Overlaps in misclassified true-*single* datasets
(iteration 300, out of 1002 true *singles*)